\documentclass[%
 reprint,
 amsmath,amssymb,
 aps,
 prd,
]{revtex4-2}
\usepackage{graphicx}
\usepackage{dcolumn}
\usepackage{bm}
\usepackage[T1]{fontenc}
\usepackage{slashed}
\usepackage[utf8]{inputenc}
\usepackage{xspace}
\usepackage{color}
\usepackage[normalem]{ulem}
\usepackage{array}
\usepackage{ulem,fancyvrb}
\usepackage{xcolor}
\usepackage{subfig}
\usepackage{multirow}
\usepackage{url}
\usepackage{hyperref}
\hypersetup{
colorlinks=true,
linkcolor=blue,
citecolor=blue,
}

\graphicspath{ {Plots/} }

\usepackage{lineno}

\begin{document}

\title{Higgs to charm quarks in 
vector boson fusion plus a photon}
\author{Ben Carlson}
\email{bcarlson@pitt.edu}
\author{Tao Han}
\email{than@pitt.edu}
\author{Sze Ching Iris Leung}
\email{SZL13@pitt.edu}
\affiliation{PITT PACC, Department of Physics and Astronomy, University of Pittsburgh, Pittsburgh, PA 15260, USA}
\date{\today}


\begin{abstract}
\noindent
    Experimentally probing the charm-Yukawa coupling in  the LHC experiments
is important, but very challenging due to an enormous QCD background. We study a new channel that can be used to search for the Higgs decay $H\to c\bar c$, using the vector boson fusion (VBF) mechanism with an associated photon. In addition to suppressing the QCD background, the photon gives an effective trigger handle. We discuss the trigger implications of this final state that can be utilized in ATLAS and CMS. We propose a novel search strategy for $H\to c\bar c$ in association with VBF jets and a photon, where we find a projected sensitivity of about 13 times the SM charm-Yukawa coupling at 95$\%$ $\text{CL}_s$ at High Luminosity LHC (HL-LHC). Our result is comparable and complementary to existing projections at the HL-LHC. We also discuss the implications of increasing the center of mass collision energy to 30 TeV and 100 TeV.
\end{abstract}


\begin{flushright}
PITT-PACC-2105
\end{flushright}

\maketitle

\section{Introduction}
Since the discovery of the Higgs boson ($H$) by the ATLAS~\cite{Aad:2012tfa} and CMS~\cite{Chatrchyan:2012ufa} collaborations, determining its properties has become a high priority for the experiments at the LHC. Higgs boson couplings to weak gauge bosons are governed by the spontaneous symmetry breaking of the gauge theory, and have been well measured. However, the mass generation of fermions is a distinctive question. In the Standard Model (SM), fermion mass terms emerge from Higgs boson Yukawa interactions and thus the Yukawa couplings are proportional to the fermion mass. Therefore, it is crucial to establish the pattern of the Yukawa couplings to fermions in order to verify the SM and seek hints of Beyond-the-Standard-Model (BSM) physics. To date, Higgs couplings to third generation fermions have been observed to $b\bar{b}$~\cite{Sirunyan:2018kst,Aad:2020jym}, $t\bar t$~\cite{Aaboud:2018urx,Sirunyan:2018hoz} and to $\tau^{+}\tau^{-}$~\cite{Khachatryan:2016vau,Aaboud:2018pen,Sirunyan:2017khh}. Direct observations of the Higgs couplings to the second generation of fermions are thus of critical importance to further confirm the non-universal pattern of Yukawa couplings~\cite{Egana-Ugrinovic:2019dqu,Bar-Shalom:2018rjs}. 
Because of its distinctive experimental signature, the decay mode $H\to \mu^+\mu^-$ via the $gg$ fusion production \cite{Han:2002gp} 
and the vector-boson fusion production (VBF) \cite{Plehn:2001qg} is promising to be observed in the near future  \cite{ATL-PHYS-PUB-2018-054,CMS-PAS-FTR-18-011}. 
Testing the Higgs Yukawa coupling to the charm quark ($y_c$), on the other hand, is known to be challenging at hadron colliders due to the formidable QCD backgrounds. 

At the LHC, billions of $pp$ collisions happen every second but only a small fraction of these events will be recorded due to limitations in data storage capacity and rate limitations of the detector readout electronics. The judicious selection
of events as using triggers such as isolated leptons, photons, or jets with high transverse momentum, are needed to record events of physical interest. Searching for the decay mode $H\to c\bar c$ with limited energy for the decay products requires incorporating the Higgs boson production mechanism to develop an efficient trigger strategy. 
There are currently two experimental probes of the charm-Yukawa coupling at the LHC. One approach is to use Higgs association production with a leptonically decaying $Z$ boson ($ZH$ channel)~\cite{Alves:2019ppy,Aaboud:2018fhh}. The $ZH$ channel provides a bound on $\mu$ of 110 for 36.1 $\text{fb}^{-1}$ of data, where $\mu$ is defined as the ratio of the new physics cross section  and the SM expectation. An extrapolation of the ATLAS analysis leads to a projection of $\mu < $ 6 using 3 $\text{ab}^{-1}$ at the HL-LHC~\cite{ATLAS:2018tmw}. A recent preliminary result from ATLAS improves the sensitivity to $\mu <$ 26 by utilizing 139$\text{fb}^{-1}$ and the leptonic decays of the $W$ boson as well as $Z$ to invisible decays~\cite{ATLAS-CONF-2021-021}. A similar result from CMS also incorporating associated production with a $W$ boson, $Z$ to invisible decay modes and utilizing substructure techniques yields an observed constraint of $\mu < $ 70~\cite{Sirunyan:2019qia}. The LHCb experiment has provided limits using 1.98 $\text{fb}^{-1}$ of data, providing an observed constraint of $\mu <$ 6400 ~\cite{LHCb-CONF-2016-006}, with a projection of an upper limit on the $\mu$ of 50
after collecting 300 $\text{fb}^{-1}$ of data at 14 TeV assuming no improvements in the detector performance or analysis \cite{Bediaga:2018lhg}.

Another approach does not rely on tagging charm quarks from the Higgs decay, but instead uses the decay of a Higgs boson $H\to J\slash\psi + \gamma$ \cite{Bodwin,Bodwin:2016edd,Bodwin:2014bpa,Perez:2015lra,Perez:2015aoa}, a process that has been searched for by ATLAS~\cite{Aad:2015sda,Aaboud:2018txb} and CMS~\cite{Sirunyan:2018fmm}. This process gives a looser bound on charm-Yukawa coupling of 50 times the SM prediction even at the HL-LHC~\cite{Aad:2015eyf}, due to the contamination from $H\to \gamma\gamma^*$ with a vector meson dominance in $\gamma^*$-$J/\psi$ mixing, which is about an order of magnitude larger than that from the direct $Hc\bar c$ coupling~\cite{Bodwin,Bodwin:2014bpa,Koenig:2015pha}. 

In Table  \ref{tab:summaryofsearches}, we collect the current results from the LHC searches (upper two rows) and the HL-LHC projection (lower two rows). For the $VH$ channel, the HL-LHC projections at CMS are estimated by simply scaling for the signal strength $\mu$ from the increase of the luminosity. 
We further translate the results to estimate the sensitivity to the modification from the SM coupling $\kappa_c = y_c^{\rm BSM} /y_c^{\rm SM}$ as described in Sec.~\ref{sec:HL-LHC}. 
For the channel $H\to J\slash\psi + \gamma$, $\kappa_c$ does not have a simple relation with $\mu$ due to the contamination from $H\to \gamma\gamma^*$ as noted above. It is instead estimated using Eq.~(53a) in~\cite{Bodwin:2014bpa}.

\begin{table*}[tbh]
\centering
\begin{tabular}{c|c|c|c|c}
\hline
Channel & \multicolumn{3}{c|}{$\sigma(VH)\times\mathcal{B}(H\rightarrow c\bar{c})$} & $\mathcal{B}(H\rightarrow J/\psi \gamma)$\\ \hline
Experiment & ATLAS& CMS& LHCb& ATLAS\\ \hline
Data set & 13 TeV & 
13 TeV & 8 TeV & 13 TeV \\ 
& 139 $\text{fb}^{-1}$ & 
35.9 $\text{fb}^{-1}$ & 1.98 $\text{fb}^{-1}$ & 36.1 $\text{fb}^{-1}$\\ \hline
$\mu$  (95$\%$ $\text{CL}_s$) & 26 \cite{ATLAS-CONF-2021-021} & 70 \cite{Sirunyan:2019qia} & 6400 \cite{LHCb-CONF-2016-006} & 120 \cite{Aaboud:2018txb}\\ \hline\hline 
HL-LHC on $\mu$ & 6.3 \cite{ATLAS:2018tmw} & 7.7$^{*}$ & 50 (300 $\text{fb}^{-1}$) \cite{Cepeda:2019klc} & 15 \cite{Aad:2015eyf}\\ \hline
HL-LHC on $\kappa_c$ & 2.7$^{\dagger}$ & 3.1$^{\dagger}$ & 7 (300 $\text{fb}^{-1}$) \cite{Bediaga:2018lhg} & 50$^{\dagger\dagger}$\\ \hline
\end{tabular}
\caption{Summary of existing search results at the LHC (upper two rows) and the HL-LHC projection (lower two rows). The CMS entry marked with $^{*}$ is scaled from the reported $\mu$ value to higher luminosity. The entries marked with $^{\dagger}$ were computed from the reported $\mu$ values (see Sec.~\ref{sec:HL-LHC}.) The entry marked with  $^{\dagger\dagger}$ is scaled according to the description in the text following \cite{Bodwin:2014bpa}. 
}
\label{tab:summaryofsearches}
\end{table*}

Rather than search for $H\rightarrow c\bar{c}$, it has also been proposed to constrain $y_c$ by requiring a charm tag in the production $gc\to Hc$ with $H\to \gamma\gamma$. This channel yields a 95$\%$ $\text{CL}_s$ limit on $\kappa_c$ ranging from 2.6 to 3.9 at HL-LHC,  depending on theoretical uncertainties  \cite{Brivio:2015fxa}. An attempt to perform a global fit for the Higgs couplings may lead to a tighter bound on the charm-Yukawa coupling
\cite{Perez:2015aoa,Coyle:2019hvs,deBlas:2019rxi,Carpenter:2016mwd}, with a few model-dependent assumptions. In particular, a 95$\%$ $\text{CL}_s$ upper bound on $\kappa_c$ of 1.2 at HL-LHC is claimed in~\cite{deBlas:2019rxi}, obtained from the upper limit of branching ratio of Higgs boson decays to untagged BSM particles assuming $|\kappa_V| \leq 1$.  Exploiting kinematic information of the Higgs boson, such as transverse momentum distribution and rapidity distribution, was proposed in probing the light-quark Yukawa couplings \cite{Soreq_2016,Cohen:2017rsk,Bishara:2016jga} and implemented in a combined fit by CMS~\cite{Sirunyan:2018sgc}. The asymmetry between $W^{+}$ and $W^{-}$ production has also been proposed to constrain  $y_c$~\cite{Coyle:2019hvs,Yu:2016rvv}. It has also been proposed to constrain $y_c$ using Di-Higgs production~\cite{Alasfar:2019pmn,Egana-Ugrinovic:2021uew}. There are proposals to further enhance the sensitivity of $H\rightarrow c\bar{c}$ by utilizing an additional photon radiation  \cite{Khanpour:2017inb,Han:2018juw,Aguilar-Saavedra:2020rgo}.

In this work, we propose a novel approach to probe the Yukawa coupling of charm quark via VBF for the Higgs boson production with an additional photon. This work builds off of the idea of introducing a new subset of the VBF production mode utilizing photon radiation as an additional handle  \cite{Mele:2007xg,Gabrielli:2007wf,Piccinini:2009zz,Asner:2010mj,Arnold:2011rm,Gabrielli:2016mdd,Biekotter:2020flu} for triggering and background suppression. The new channel we propose can provide complementary information to the existing searches using the $WH$ and $ZH$ channels.

The rest of the paper is organized as follows. We first lay out our search strategy in Sec.~\ref{sec:strategy}, in particular a proposal for triggering the signal events. We then present our analyses and the results in Sec.~\ref{sec:results}, including HL-LHC. We finally summarize our results and draw the conclusion in Sec.~\ref{sec:conclude}.

\section{Proposed search strategy}
\label{sec:strategy}

\begin{figure}[h]
    \centering
    \subfloat{{\includegraphics[width=4cm]{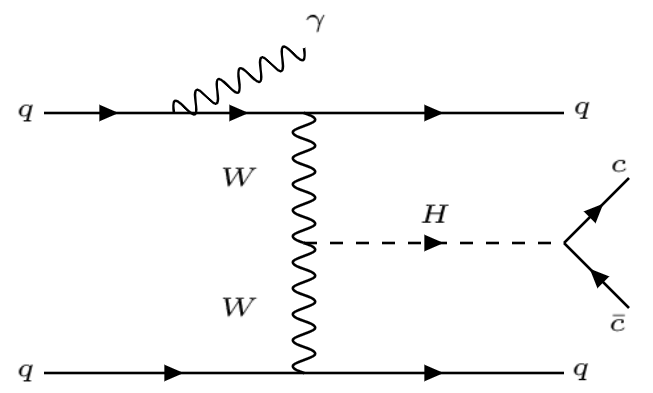} }}%
    \qquad
    \subfloat{{\includegraphics[width=3.7cm]{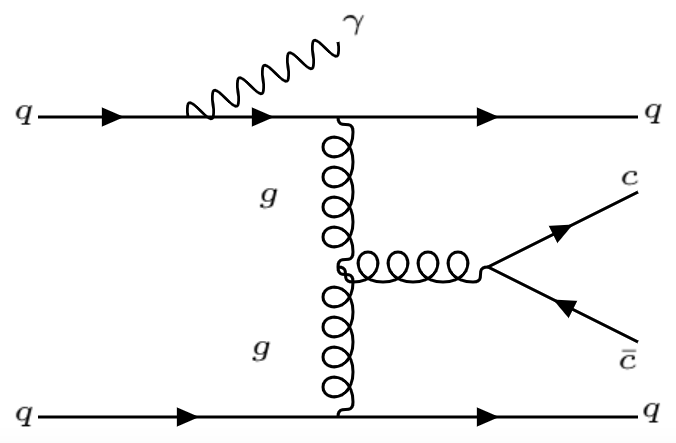} }}%
    \caption{Representative Feynman diagrams of the signal channel (left) and QCD  multi-jet background (right)}
    \label{fig:feynman}
\end{figure}

The decay branching fraction for $H\rightarrow c \bar{c}$ is about $3\%$, which leads to a cross section about 0.1 pb (1 pb) from the VBF ($gg$) production at the 13 TeV LHC \cite{deFlorian:2016spz}. This yields a sizeable signal sample with the currently achievable luminosity. However, the process $H\rightarrow c \bar{c}$ is not only difficult to trigger, but also challenging to distinguish from large QCD multi-jet background. More sophisticated search strategies should be developed to reach the needed sensitivity for signal observation. First, the VBF channel has striking experimental signature where a central Higgs boson is accompanied by two light jets with a large rapidity gap. Second, the addition of the photon improves the trigger efficiency compared to what can be achieved using only multi-jet final states as well as suppresses the gluon-rich dominant multi-jet background. We therefore propose to search for the signal process 
\begin{equation}
p p \rightarrow qqH \gamma \ \ {\rm with}\ \ 
H \rightarrow c\bar{c}.
\label{eq:signal} 
\end{equation} 
Our signal process has distinctive features, which are characterized by two $c$-jets from the Higgs boson decay and an energetic photon in the central region, with two light jets separated by a large rapidity gap.
The dominant background is QCD multi-jet production associated with a photon, where at least two jets are tagged (or mistagged) as $c$-jets. Other backgrounds include $Z\gamma+$jets and VBF Higgs production$+\gamma$ with $H\rightarrow b \bar{b}$, where both $b$-jets are mistagged as $c$-jets. However, their contributions are expected to be much less significant than the QCD multi-jet background, thus not included in this analysis. The representative Feynman diagrams of the signal and leading background are shown in Fig.~\ref{fig:feynman} for illustration.

 Our analyses are designed to isolate the signal based on their kinematic features, described in detail in Sec.~\ref{sec:results}. To estimate the achievable sensitivity to the signal at the hadronic collider environment, we  use  simulated events from Monte Carlo tools. 
 %
 With the very large hadronic production rate at the LHC, the trigger system is designed to record events of physical interest. For the relatively soft final states in our signal process, a dedicated trigger strategy is essential to record the signal events. 

\subsection{Monte Carlo simulation}
\label{sec:MC}

Our targeted signal as seen in Eq.~(\ref{eq:signal}) is $H\to c\bar c$ plus an additional $\gamma$. 
Both signal and QCD multi-jet background are generated at LO with \textbf{MG5@MCNLO v2.6.5}~\cite{Alwall:2014hca} at the $pp$  collider center of mass (c.m.) energy $\sqrt{s}=$ 13 TeV using the 
PDF4LHC15\textunderscore nlo\textunderscore mc PDFs~\cite{Butterworth:2015oua}. The Higgs boson in the signal process is then decayed into $c\bar{c}$ by \textbf{MadSpin}~\cite{Artoisenet:2012st}. The renormalization and factorization scales are set to be at the EW scale of the $W$-mass ($m_W$). To enhance generation efficiency, both samples are generated with the following parton-level requirements, which are slightly looser than the thresholds used in analysis. We require two VBF jets inside the detector acceptance in pseudo-rapidity ($\eta$), with transverse momenta
\begin{equation}
    p_\text{T}^j > 35 \, \text{GeV},\ \ \lvert \eta_j \rvert < 5,
    \label{eq:basic1}
\end{equation}
and an isolated photon in the central region with transverse momentum
\begin{equation}
    p_\text{T}^\gamma > 25 \, \text{GeV},\ \ \lvert \eta_\gamma \rvert < 3.
        \label{eq:basic2}
\end{equation}

The parton shower and hadronization are simulated with \textbf{Pythia8}~\cite{Sjostrand:2014zea} and a fast detector simulation is implemented with \textbf{Delphes3} using the default cards~\cite{Ovyn:2009tx,deFavereau:2013fsa}. Jets are reconstructed using the anti-$k_{\text{t}}$ algorithm~\cite{Cacciari:2008gp} with a radius parameter 
\begin{equation}
R = 0.4.
        \label{eq:basic3}
\end{equation} 
After these basic acceptance cuts, 
the cross-sections of signal and leading  background processes at different center-of-mass energies are listed in Table \ref{tab:crosssections} using the same calculation set-up.
We see that the signal rates are sizeable with the current and anticipated luminosities. However, the signal-to-background ratios are quite low, roughly at the order of $10^{-5}$, rendering the signal identification extremely challenging.

\begin{table}[h]
\centering
\begin{tabular}{c|c|c|c|c}
\hline
 & 13 TeV & 14 TeV & 30 TeV & 100 TeV\\ \hline
$\sigma_{\text{VBF}+\gamma}$ (pb) & 0.024 & 0.027 & 0.099 & 0.43\\ \hline
$\sigma_{p p \rightarrow 4j + \gamma}$ (pb) & 830 & 940 & 3700 & 21000\\ \hline
\end{tabular}
\caption{Cross sections of signal and background at different center-of-mass energies, with the basic acceptance cuts in Sec.~\ref{sec:MC}.}
\label{tab:crosssections}
\end{table}

\subsection{Proposed trigger strategy}
The ATLAS~\cite{Aaboud:2016leb} and CMS~\cite{CMS-TRG-12-001} experiments both contain a two-level trigger system. The first level, or level-1, trigger system is composed of custom electronics while the
second level, or high-level trigger (HLT) runs software algorithms. The level-1 trigger systems of both experiments currently only utilize information from the calorimeter or muon subs-systems, and as a consequence, have shared items to trigger on electrons and photons~\cite{Aad:2019wsl}. Furthermore, due to the importance of selecting 
electrons with $p_{T}\approx m_{W}/2$, approximately 25\% of the total level-1 rate is devoted to electron/photon triggers. More information on the breakdown of rate and specific implementation can be found in the documents describing the ATLAS trigger menu ~\cite{ATL-DAQ-PUB-2016-001,ATL-DAQ-PUB-2017-001,ATL-DAQ-PUB-2018-002,ATL-DAQ-PUB-2019-001}. 

After events have been selected from the level-1 trigger using a single EM object, these events can be used to seed a variety of triggers in the HLT, including requiring additional VBF jets or jets with a $b$-tag to reduce the rate. Relevant to our current considerations, the ATLAS analyses described in \cite{Aaboud:2018gay,Aad:2020vbr} utilize a trigger with the following offline requirements: 
\begin{itemize}
    \item Photon $E_\text{T}^\gamma >$  30 GeV;
    \item At least four jets with $p_\text{T}^j >$  40 GeV;
    \item At least one pair of jets with $m_{jj} >$ 700 GeV;
    \item At least one $b$-tagged jet with 77\% efficiency. 
\end{itemize}

There are also VBF triggers described in~\cite{ATL-DAQ-PUB-2019-001,ATLAS:2021tnq} with higher $m_{jj}$ threshold and jet $p_{T}$ threshold, which would have a lower acceptance for the genuine VBF events. To develop a trigger for $H \rightarrow c\bar{c}$ final states, it seems plausible that the trigger described above for $H\rightarrow b\bar{b}$ could be modified for charm final states, by either
requiring a charm tag or by raising the $m_{jj}$ threshold. We leave the exact details and optimization to the experiments, and proceed by providing motivation for the use case of such a trigger. 

\section{Analyses and results}
\label{sec:results}

\subsection{Cut-based analysis}
\label{sec:cut-based}

To obtain a physical intuition on the characteristics of our signal and background processes, we start with a simple cut-based analysis which utilize thresholds on different kinematic observables.

\vskip 0.2cm
\noindent
\textbf{Pre-selections} \\ 
The event pre-selections, inspired by \cite{Aaboud:2018gay,Aad:2020vbr}, aim at capturing the basic features of a VBF signal. We require a photon in the central region with
\begin{equation}
p_\text{T}^\gamma > 30~{\rm GeV},\quad \lvert \eta^\gamma \rvert < 1.37\ \ {\rm or}\ \  1.52 < \lvert \eta^\gamma \rvert < 2.37. 
\end{equation}
Events are required to have 4 jets with 
\begin{equation}
p^j_\text{T} >  40~{\rm GeV\ \  and\ \ } \lvert \eta^j \rvert < 4.4. 
\end{equation}
At least 2 jets in the central region $\lvert \eta^j \rvert < 2.5$ are required to be $c$-tagged using fixed efficiency values depending on the truth flavor of the jet inspired by~\cite{Aaboud:2018fhh}. Charm jets are assumed to have a tagging  efficiency of 41$\%$, while $b$-jets have a contamination probability of 25\% and light jets a contamination probability of 5$\%$. The two highest-$p_\text{T}$ $c$-tagged jets are identified as signal jets from Higgs boson decay while the remaining jets are identified as the VBF jets. The VBF jet pair is required to have invariant mass of at least 800 GeV so that the trigger requirement is fully efficient. In addition, the signal $c$-jet pair is required to have $p_\text{T}^{cc} >$  80 GeV to remove potential bias in $m_{cc}$ distribution caused by the $p_\text{T}^j$ thresholds in the trigger. 

\begin{figure*}[th!]%
    \centering
    \includegraphics[width=0.45\textwidth]{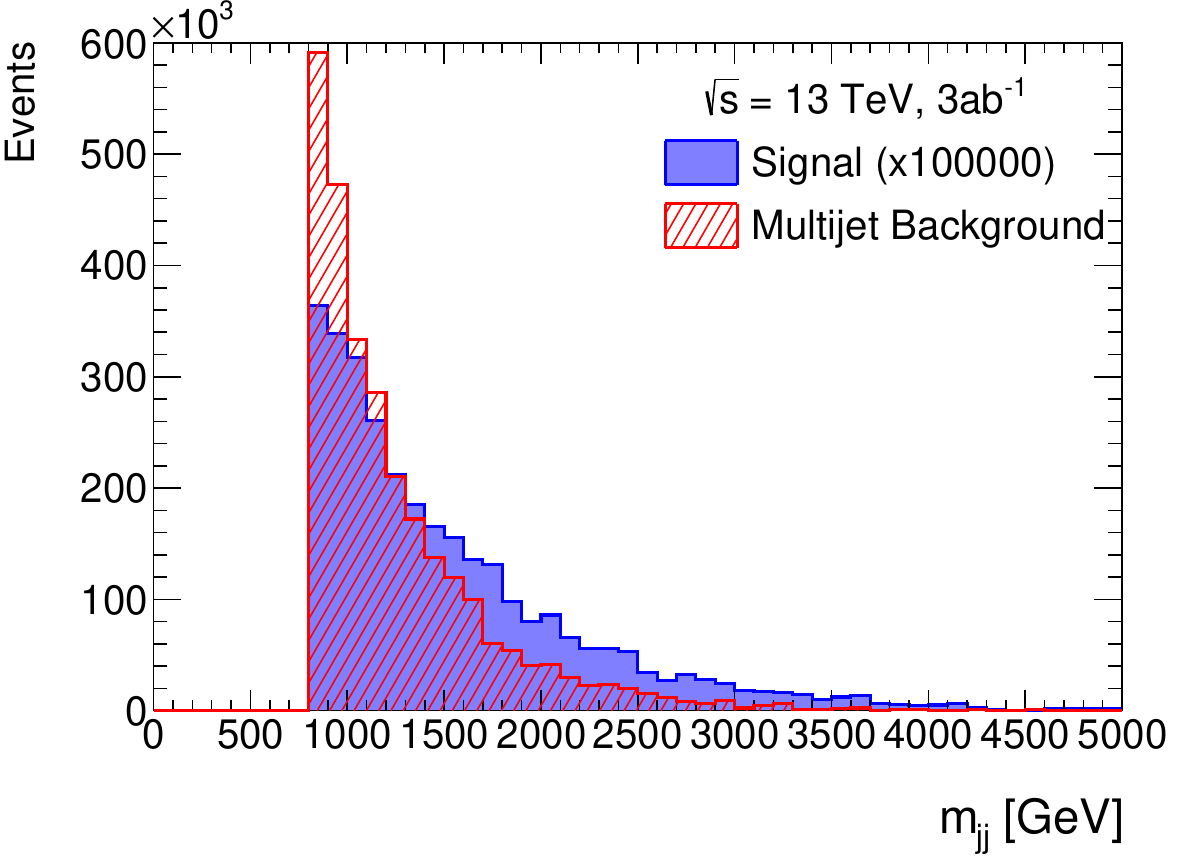} 
    \includegraphics[width=0.45\textwidth]{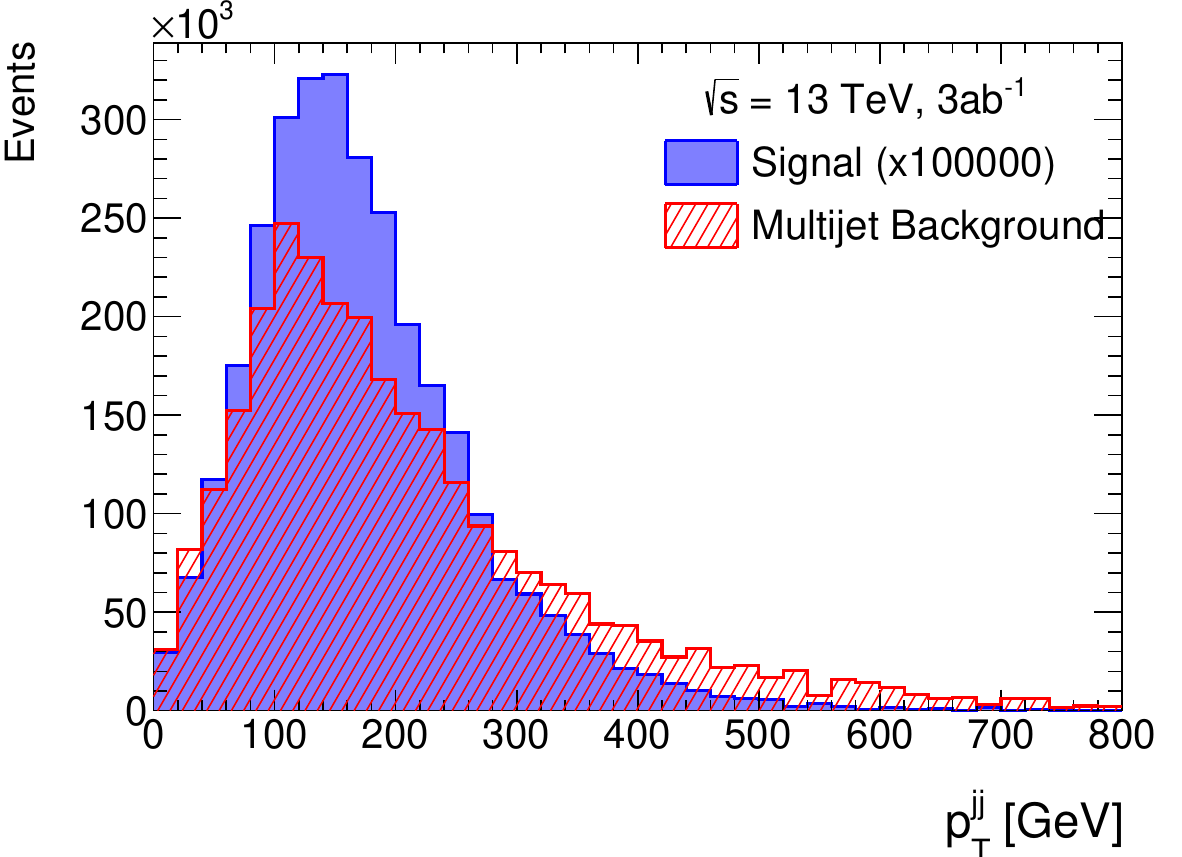}
    \includegraphics[width=0.45\textwidth]{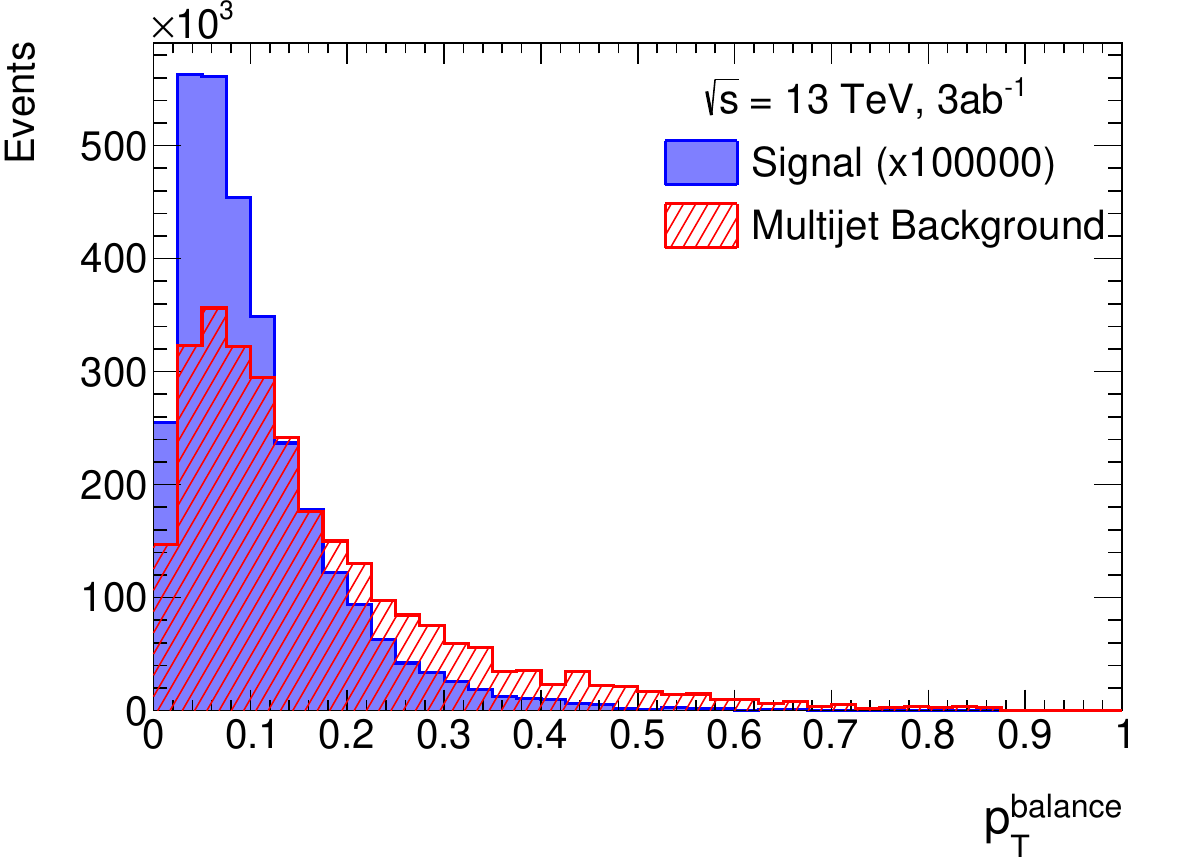}
    \includegraphics[width=0.45\textwidth]{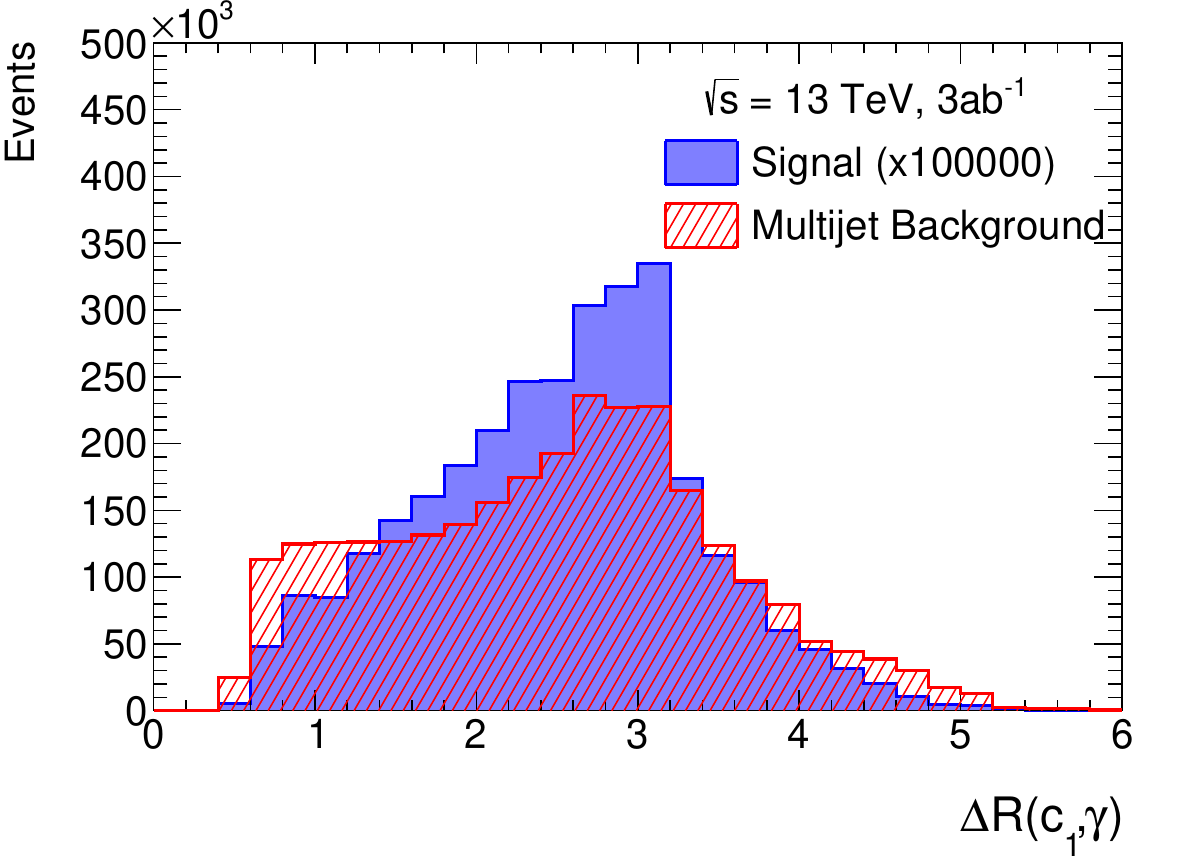}
    \includegraphics[width=0.45\textwidth]{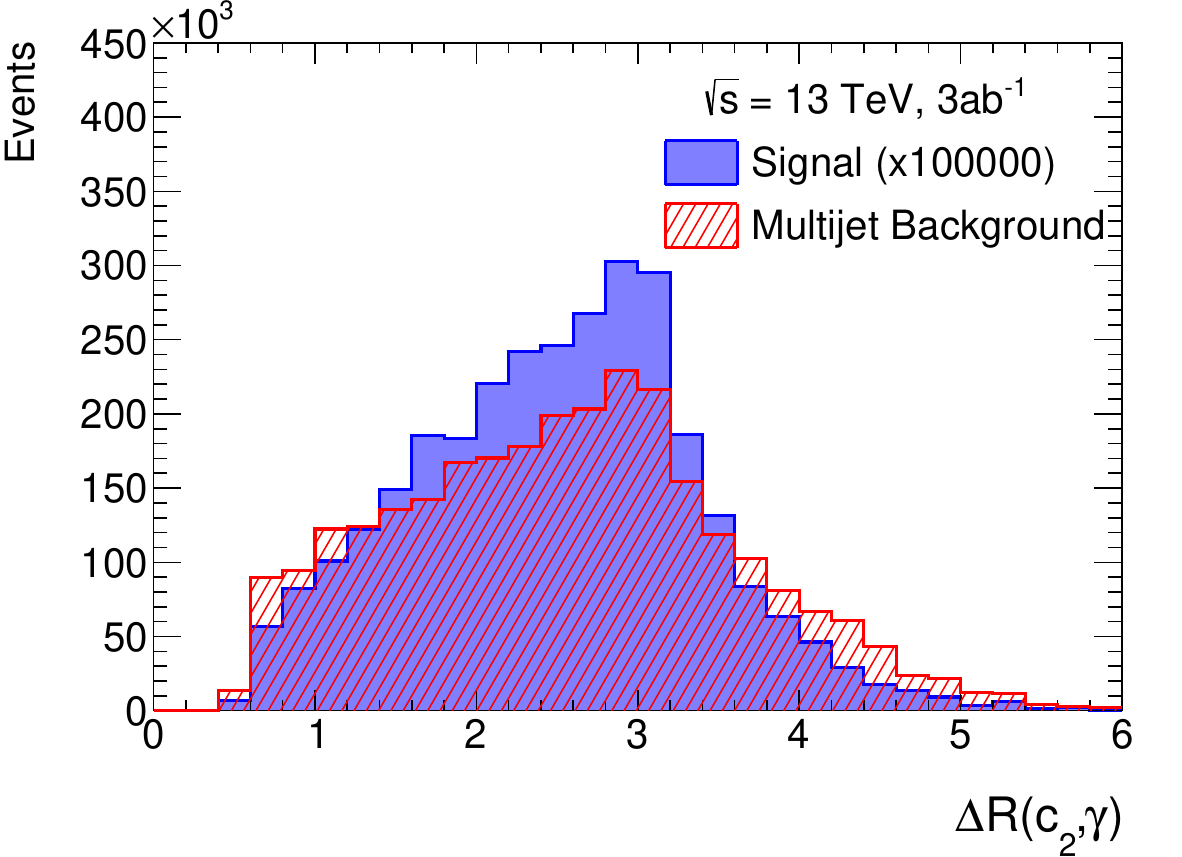}
    \includegraphics[width=0.45\textwidth]{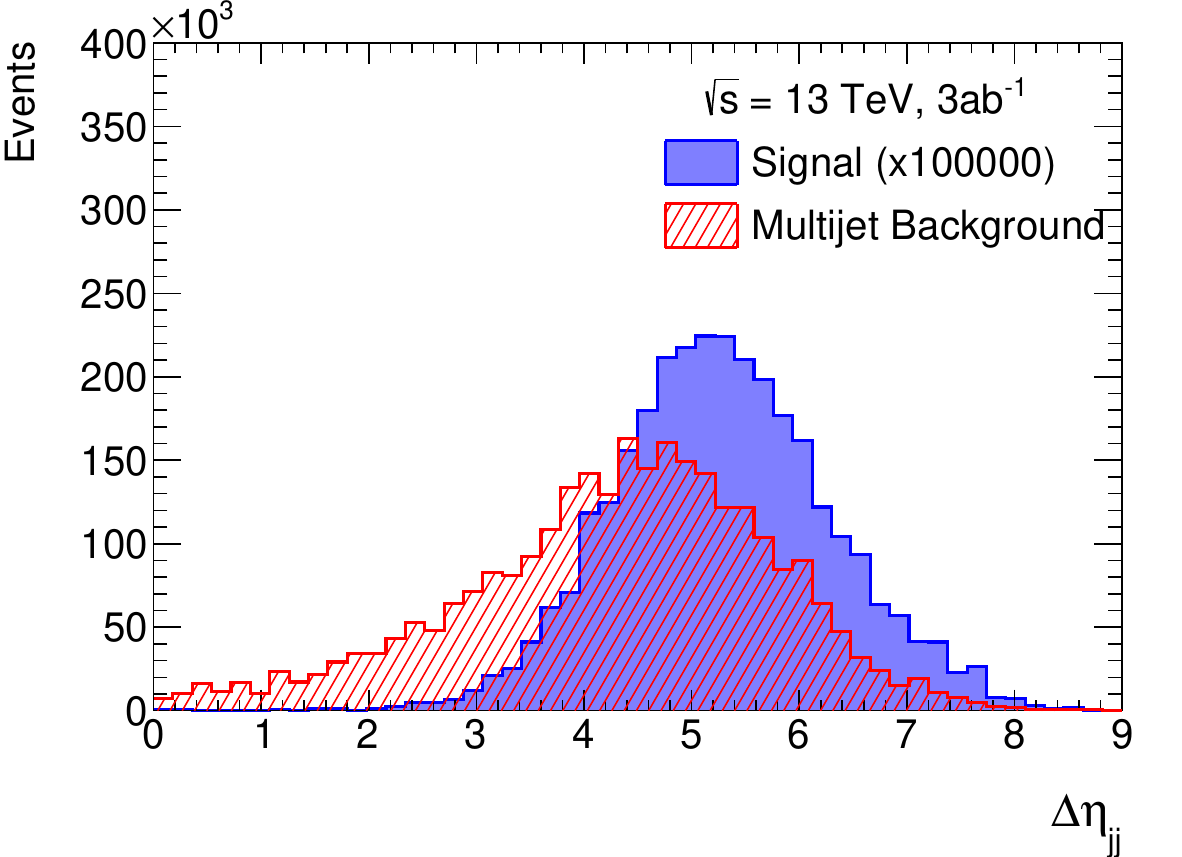}
    \caption{Distributions of useful observables for signal (blue) and multi-jet background (red) after pre-selections.}
    \label{fig:distributions1}
\end{figure*}

\begin{figure*}[th!]%
    \centering
    \includegraphics[width=0.45\textwidth]{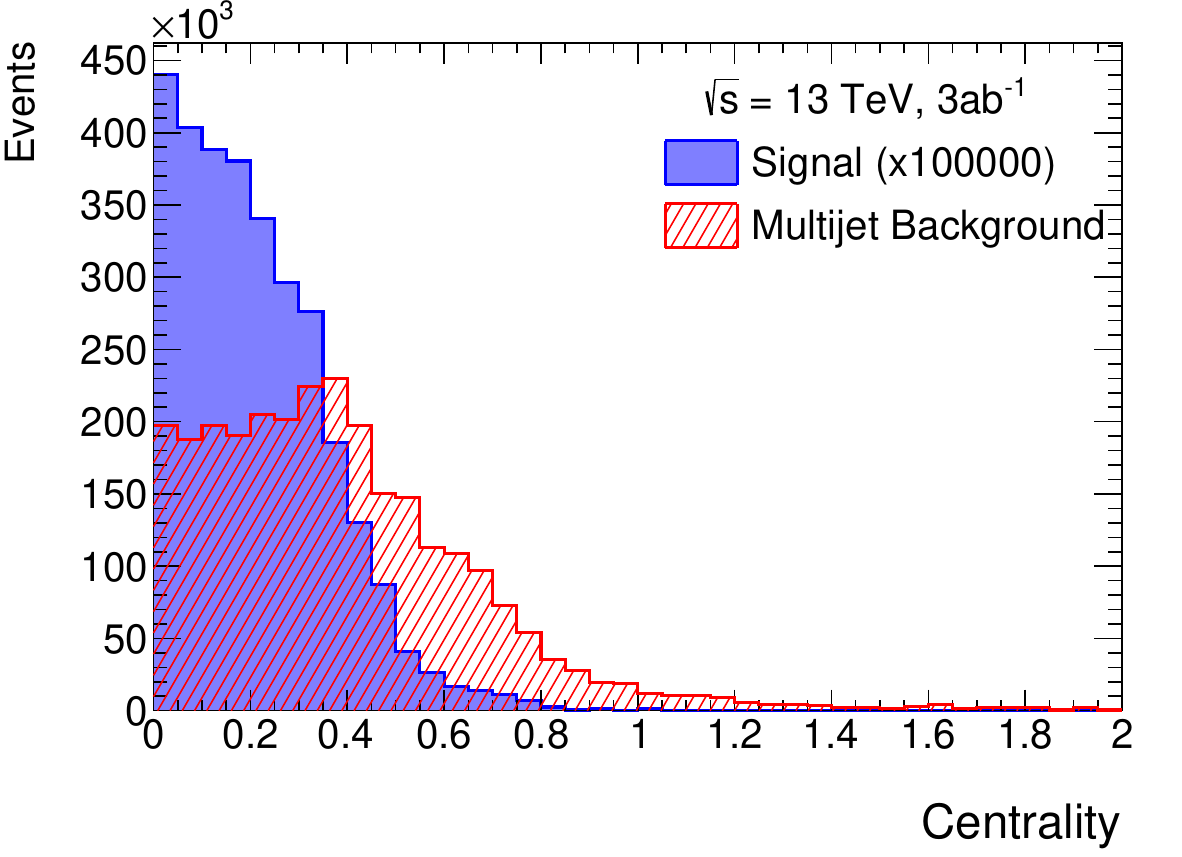}
    \includegraphics[width=0.45\textwidth]{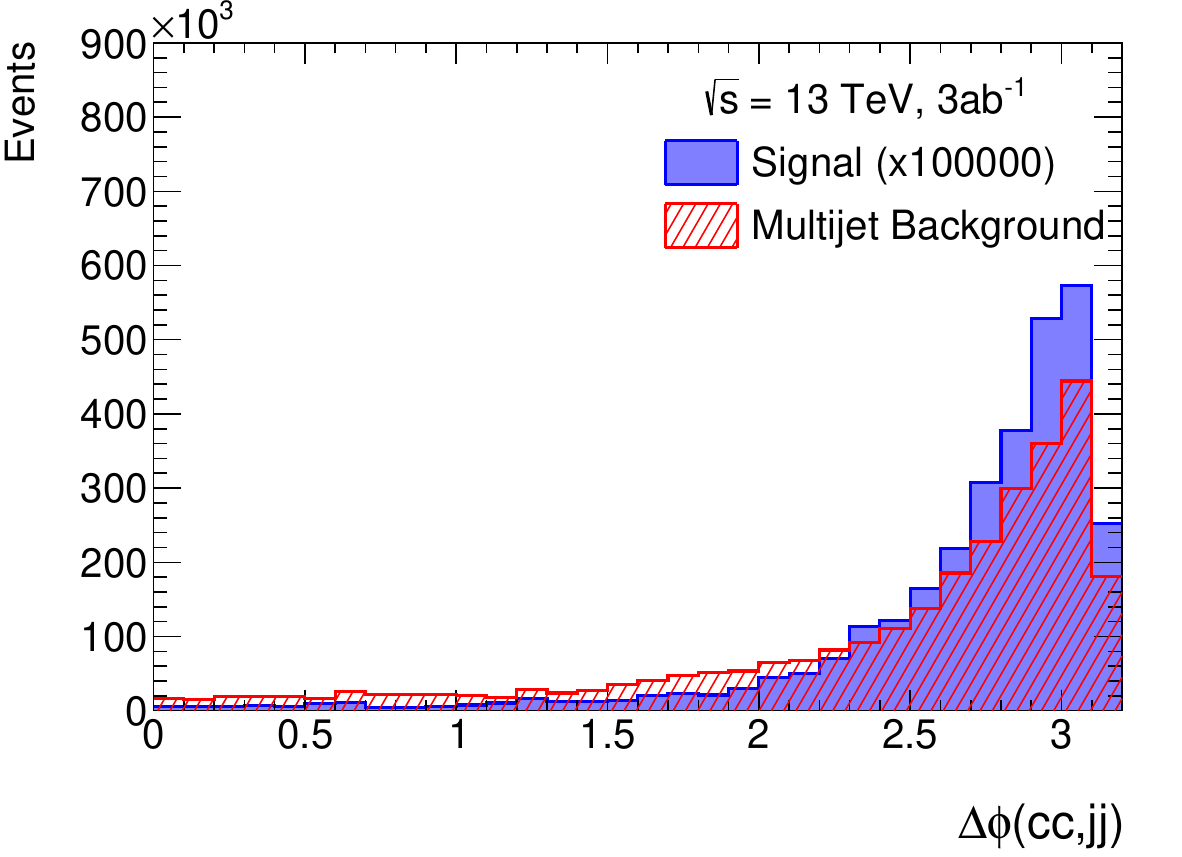}
    \includegraphics[width=0.45\textwidth]{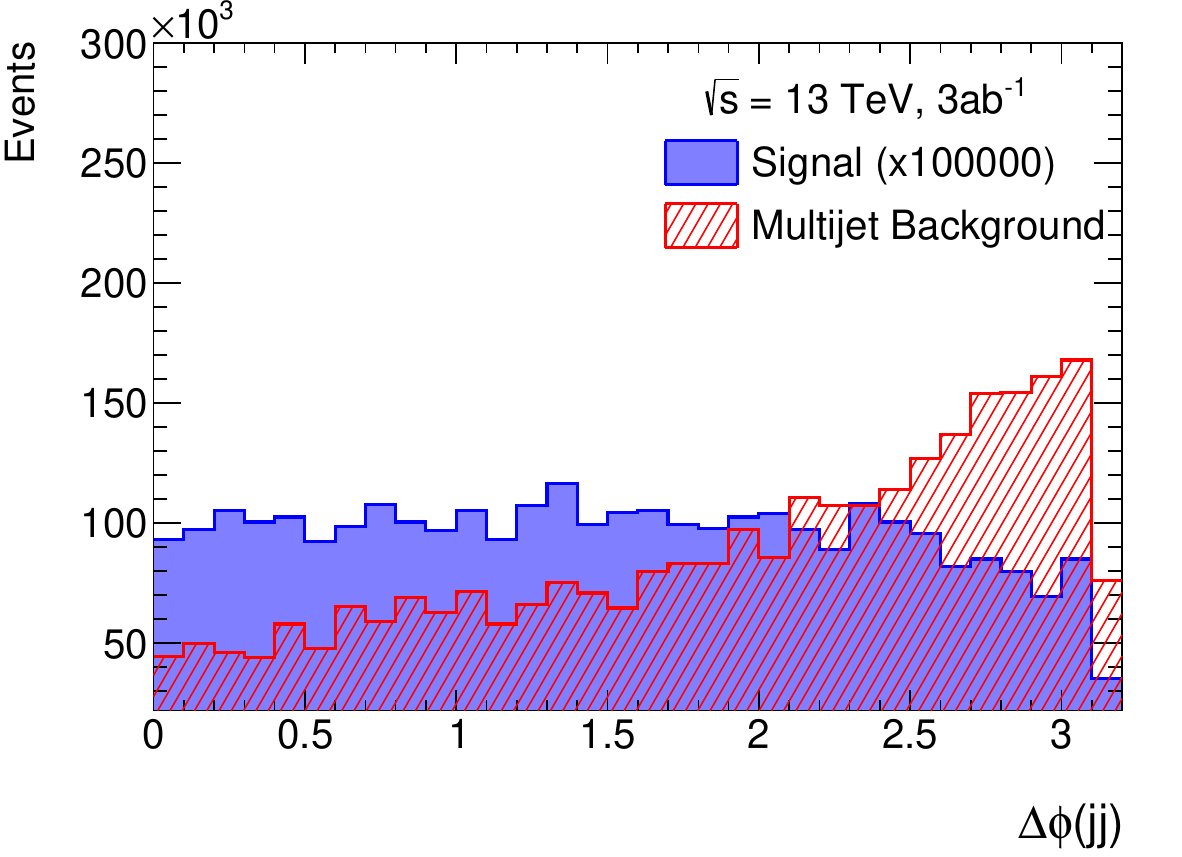}
    \includegraphics[width=0.45\textwidth]{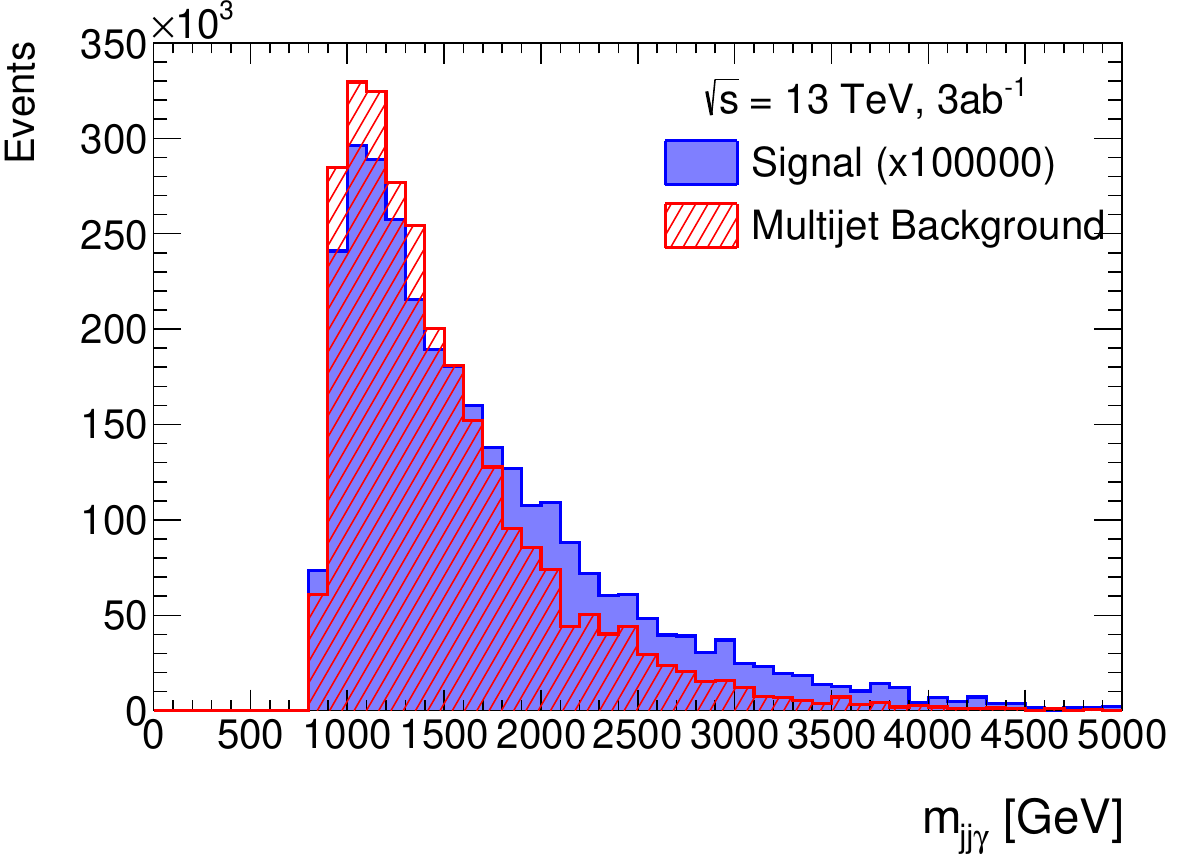}
    \includegraphics[width=0.45\textwidth]{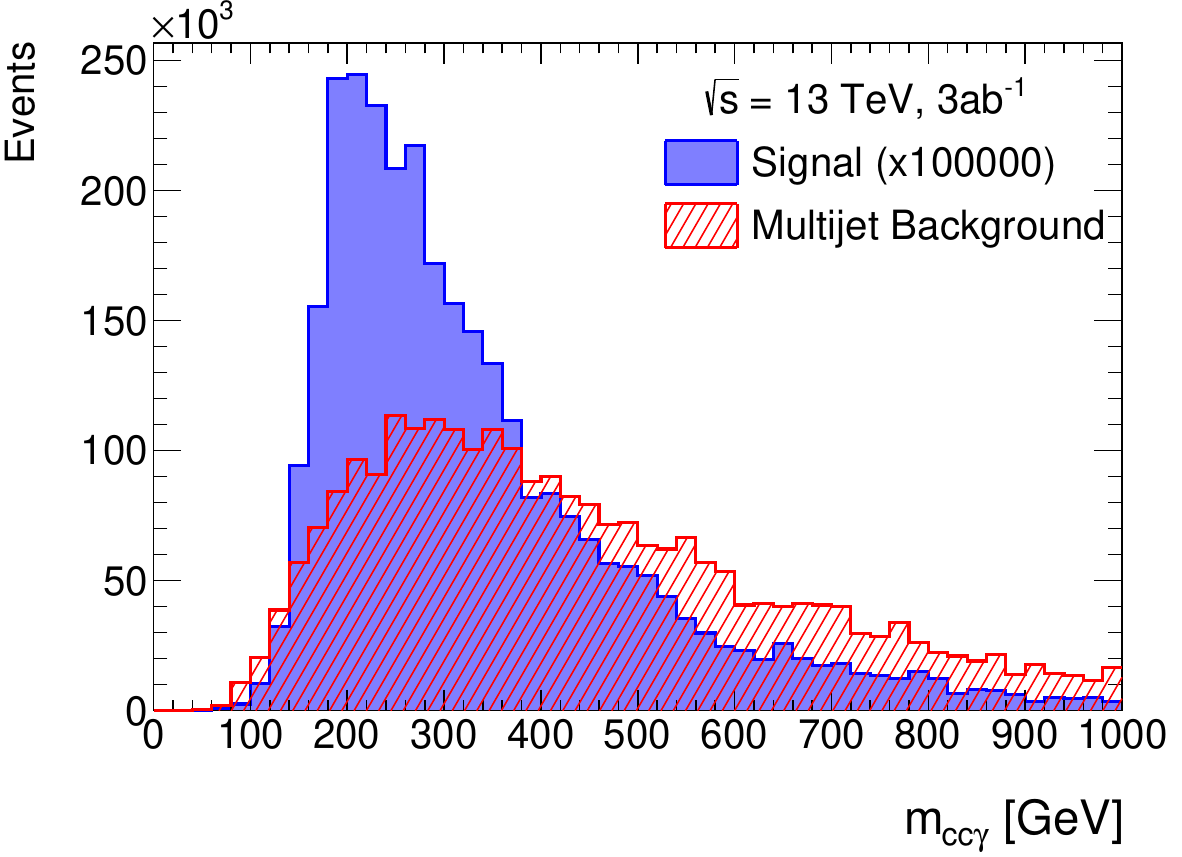}
    \caption{Distributions of useful observables for signal (blue) and multi-jet background (red) after pre-selections.}
    \label{fig:distributions2}
\end{figure*}

\vskip 0.2cm
\noindent
\textbf{Optimized analysis selections}\\
In order to exploit the full phase space of the kinematic features, we choose several additional kinematic observables which are useful to further distinguish signal and background. The distributions of these observables after the pre-selections are shown in Figure~\ref{fig:distributions1} and \ref{fig:distributions2}. 
In addition to the pre-selections, we make some further judicious cuts based on the signal kinematics as follows.
The invariant mass of the VBF jets (and photon) is large due to their back-to-back nature, so we require 
\begin{equation}
    m_{jj},\ \ m_{jj\gamma} > 1000 \, \text{GeV}.
\end{equation}
The transverse momentum of the VBF jet pair is governed by the $W/Z$ exchange and thus relatively low. We limit their value to be 
\begin{equation}
    p_\text{T}^{jj} = |\vec{p}_\text{T}^{\,j_1}
    + \vec{p}_\text{T}^{\,j_2}
    | < 300 \, \text{GeV}.
\end{equation}
Since final states from electroweak processes tend to be more back-to-back than the QCD multi-jet background, we select events with the following ratio between the magnitudes of the vector and scalar sums of the jets and photon momenta  
\begin{equation}
    p_\text{T}^\text{balance} = \frac{|\vec{p}_\text{T}^{\,j1} + \vec{p}_\text{T}^{\,j2} + \vec{p}_\text{T}^{\,c1} + \vec{p}_\text{T}^{\,c2} + \vec{p}_\text{T}^{\,\gamma} |}{p_\text{T}^{j1} + p_\text{T}^{j2} + p_\text{T}^{c1} + p_\text{T}^{c2} + p_\text{T}^{\gamma}} < 0.2
\end{equation}
Furthermore, because VBF signal features a large rapidity gap between the two forward-backward jets, events with large  pseudo-rapidity separation between the two jets are selected for 
\begin{equation}
    \Delta\eta_{jj} = |\eta_{j_1} - \eta_{j_2}| > 4 .
\end{equation}
The reason that the multi-jet background also peaks at a relatively high value in Figure~\ref{fig:distributions1} is due to the $m_{jj}$ requirement in the  pre-selections. 

As the photon is not radiated from $c$-jets from Higgs decay in our signal process, the angular separation between the signal $c$-jets and the photon tends to be larger in contrast to the QCD processes. Therefore, we require
\begin{equation}
    \Delta R(c_{1,2}, \gamma) > 1.4,
\end{equation}
where $c_{1,2}$ are leading and sub-leading $c$-jets. We also make use of the centrality of the photon relative to the VBF jets and require 
\begin{equation}
    \text{centrality} = \lvert \frac{y_\gamma-\frac{y_{j1}+y_{j2}}{2}}{y_{j1}-y_{j2}} \rvert < 0.35,
\end{equation}
where $y$ is the rapidity of the jet or photon.\footnote{For a massless object, the rapidity $y$  reduces to the pseudo-rapidity $\eta$.}
Additionally, we utilize the azimuthal   angular information on the transverse plane and require
\begin{equation}
    \Delta \phi (cc,jj) > 2.3, \ \ \Delta\phi(jj) < 2.1
\end{equation}
It should be noted that the $\Delta \phi (jj)$, which is motivated by~\cite{Eboli:2000ze}, has not been used to our knowledge before in the $H\rightarrow b\bar{b}$ searches. We then require the invariant mass of the $c$-jets and photon system
\begin{equation}
    m_{cc\gamma} < 700 \, \text{GeV}. 
\end{equation}
Since this observable is highly correlated with $m_{cc}$, we choose a relatively loose cut here.

\begin{figure}[h]
    \centering
    \includegraphics[width=0.45\textwidth]{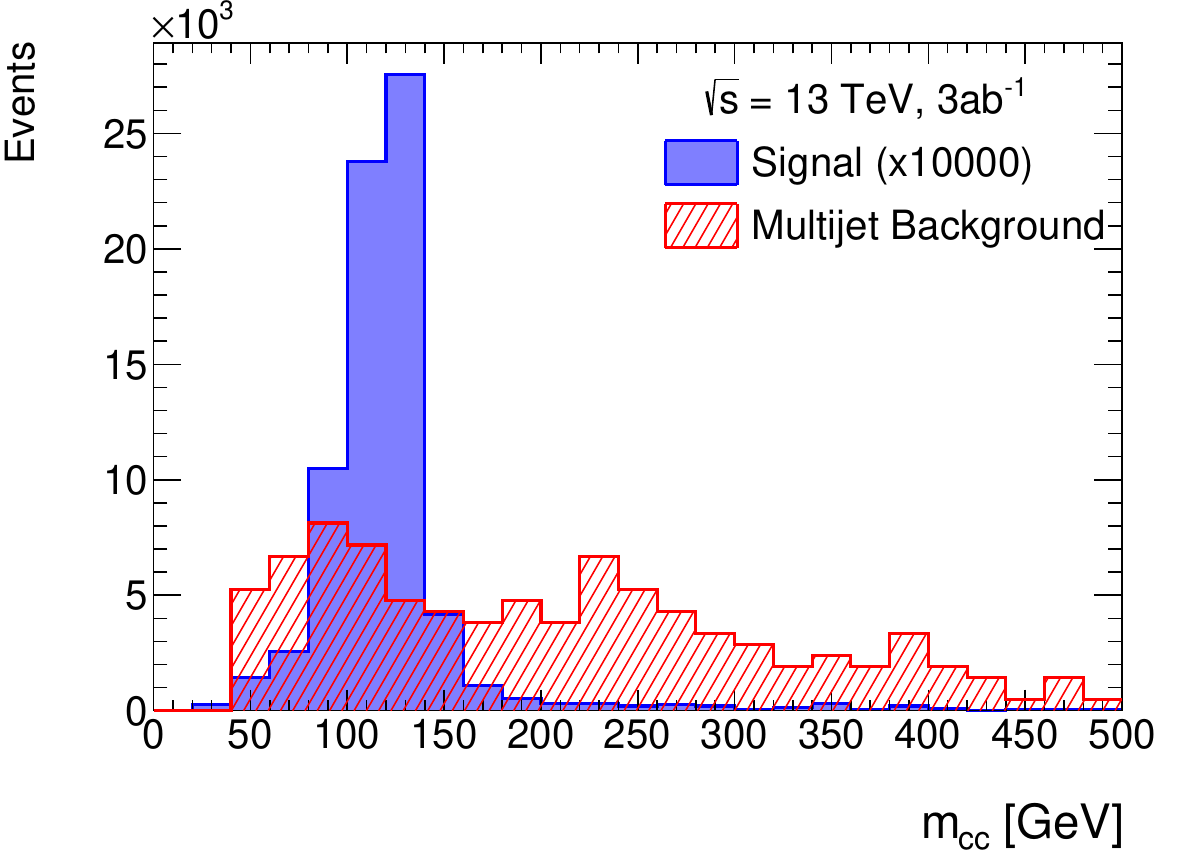}
    \caption{Distribution of the invariant mass of the signal $c$-jet pair after cuts.}
    \label{fig:masscut}
\end{figure}

The invariant mass of the signal $c$-jet pair $m_{cc}$ is used as the final discriminant. After the above selection requirements, the distribution of $m_{cc}$ is shown in Fig.~\ref{fig:masscut}. 
As expected, the Higgs signal peaks at $m_H\approx 125$ GeV, while the multi-jet  background has a more flat shape. Anticipating finite jet-mass resolution, we therefore require the $c$-jet pair to be in the Higgs mass window 
\begin{equation}
    100\ \text{GeV} < m_{cc} < 140\  \text{GeV}.
    \label{eq:mass}
\end{equation}
The expected numbers of events from signal and background of integrated luminosity 3 $\text{ab}^{-1}$ are shown in Table \ref{tab:numofevents}. The truth flavor components of the $c$-tagged jet pairs in background are also shown in Table \ref{tab:numofevents}. The major component is true $c$-jet pair as expected, but it should also be noted that the sub-leading components mainly involve light jets mistagged as $c$-jet. This suggests that improving the discrimination between $c$-jets and light jets can enhance the significance of such search.

\begin{table*}[th!]
\centering
\begin{tabular}{c|c|c}
\hline
\multicolumn{3}{c}{Cut-Based} \\ \hline
& S & B\\ \hline
\multirow{2}{*}{Pre-selections} & \multirow{2}{*}{31} & $2.8\times10^6$\\
& & ($cc$:\ 38$\%$, $cb$:\ 5.6$\%$, $cj$:\ 28$\%$, $bb$:\ 4.6$\%$, $bj$:\ 5.7$\%$, $jj$:\ 18$\%$) \\ \hline
Optimized selections & 7.4 & $8.8\times10^4$ \\ \hline
mass cut Eq.~(\ref{eq:mass})
& 5.1 & $1.2\times10^4$ \\ \hline \hline
$S/\sqrt{B}$&\multicolumn{2}{c}{0.047} \\ \hline
\end{tabular}
\caption{Expected yields of signal and multi-jet background at the HL-LHC with 3 $\text{ab}^{-1}$ from a cut-based analysis.}
\label{tab:numofevents}
\end{table*}

\subsection{Multivariate analysis}
\label{sec:multi}

Recent analyses of data in high energy physics have made extensive use of machine learning techniques, including the use of boosted decision trees (BDT) \cite{Abazov:2006gd,Sirunyan:2018kst, Aad:2020jym}. 
To improve the sensitivity reached by the simple cut-based studies in the last section \ref{sec:cut-based}, a multivariate analysis is employed, starting with only the pre-selection cuts. A BDT is trained using the \textbf{TMVA}~\cite{Hocker:2007ht} package with the same set of observables shown in Figures~\ref{fig:distributions1} and \ref{fig:distributions2} as inputs.

\begin{figure}
    \centering
    \includegraphics[width=0.45\textwidth]{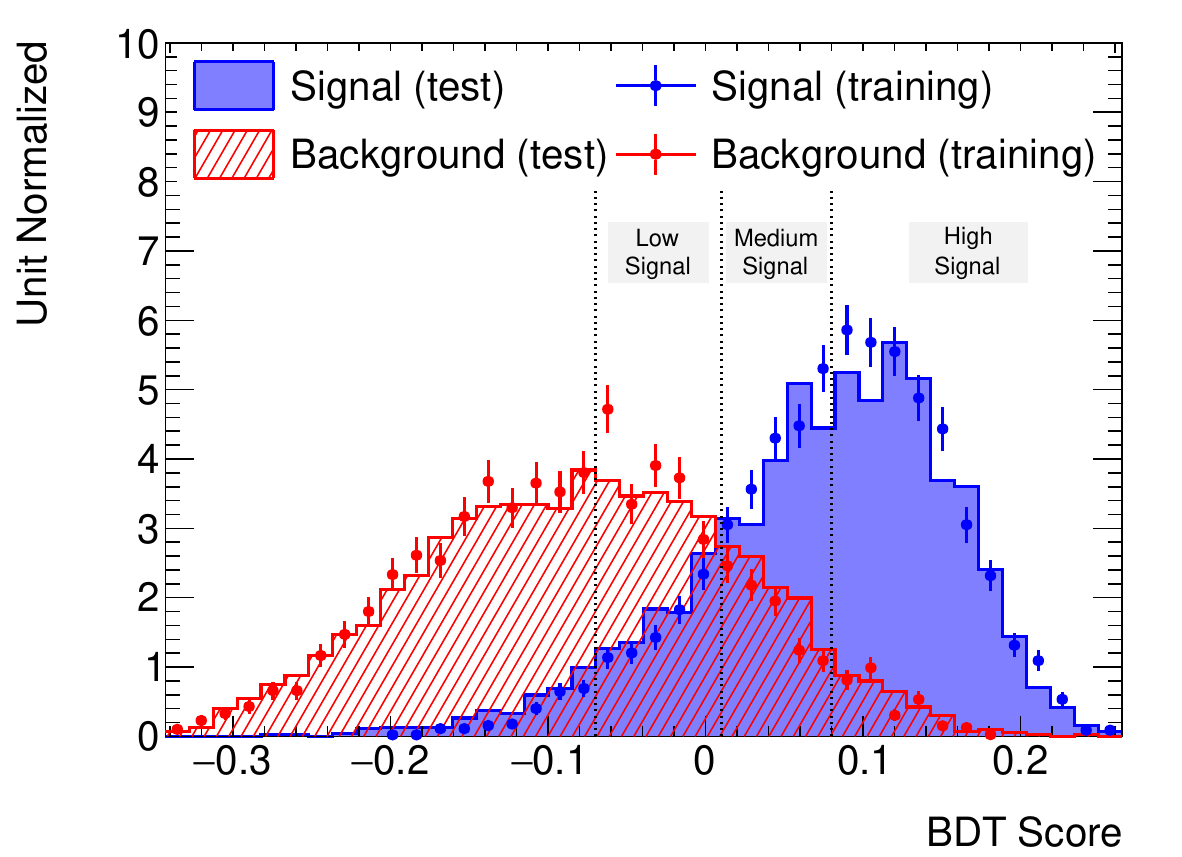}
    \includegraphics[width=0.45\textwidth]{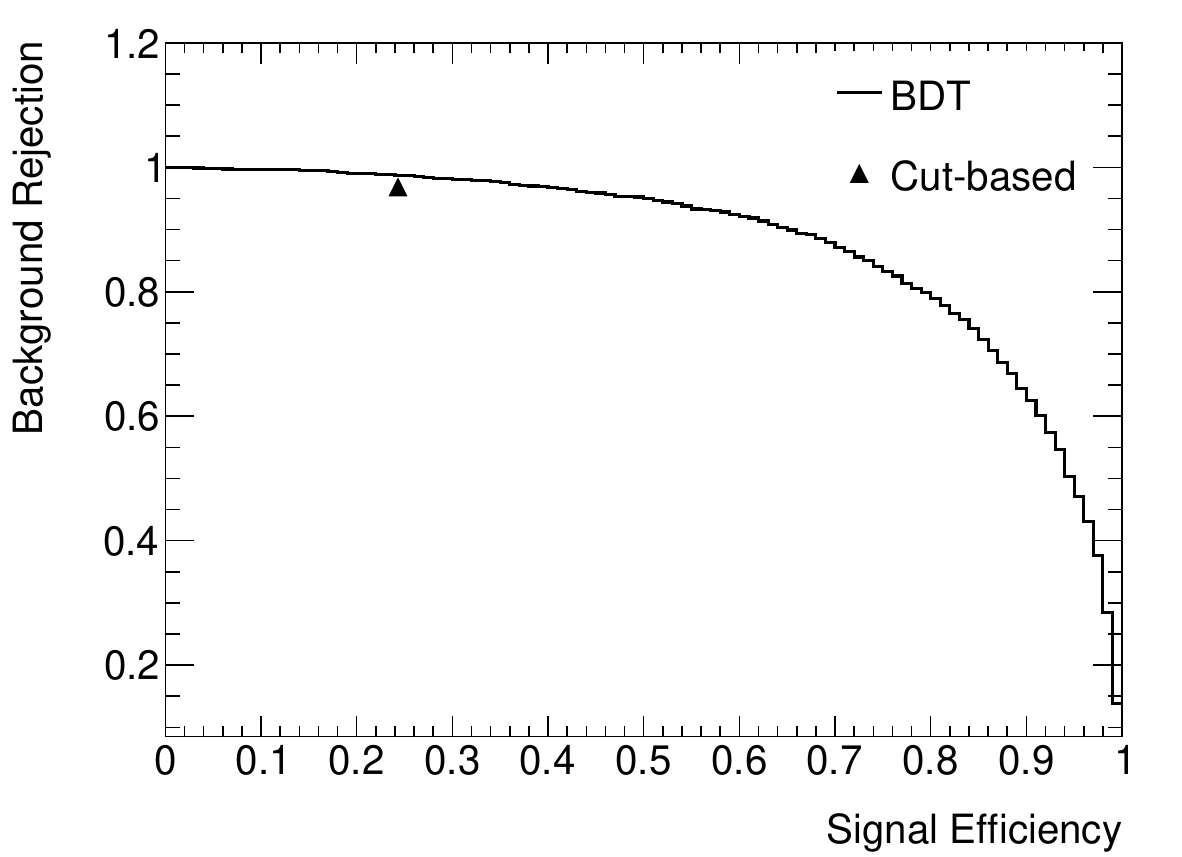}
    \caption{The distribution of the BDT score with low, medium and high signal region defined as $-$0.07-0.01, 0.01-0.08 and $>$ 0.08 respectively (top) and receiver operating characteristic (ROC) curve of the BDT (bottom). }
    \label{fig:BDT}
\end{figure}

The BDT is constructed by 850 trees, each with maximum depth of 3. A small depth is chosen because they are less susceptible to over-training but still perform very well with the aid of boosting algorithms. At each node of a tree, events are split into two subsets by cutting on an observable. The performance of the separation is assessed by Gini Index, defined by $p(1-p)$ where $p$ is the ratio of signal events to all events in that node. Therefore, a pure signal or background node corresponds to a zero Gini Index. The event sample is randomly split by half into training and test samples, where the former is used for BDT training and the latter is used for analysis and deriving limits. 
The distribution of the BDT score from signal and background along with a receiver operator characteristic (ROC) curve are shown in Figure~\ref{fig:BDT}. 
The background rejection in the ROC curve on the right panel is defined as one minus the background survival probability after the selection cuts. The BDT performs as expected with a positive score as more ``signal-like'' and a negative score as more ``background-like''. We can see that the separation between signal and background is fairly well. The test sample distribution is superimposed with the training sample distribution, showing similar performance and thus indicates no occurrence of over-training. 

\begin{figure*}[th!]
    \centering
    \includegraphics[width=0.32\textwidth]{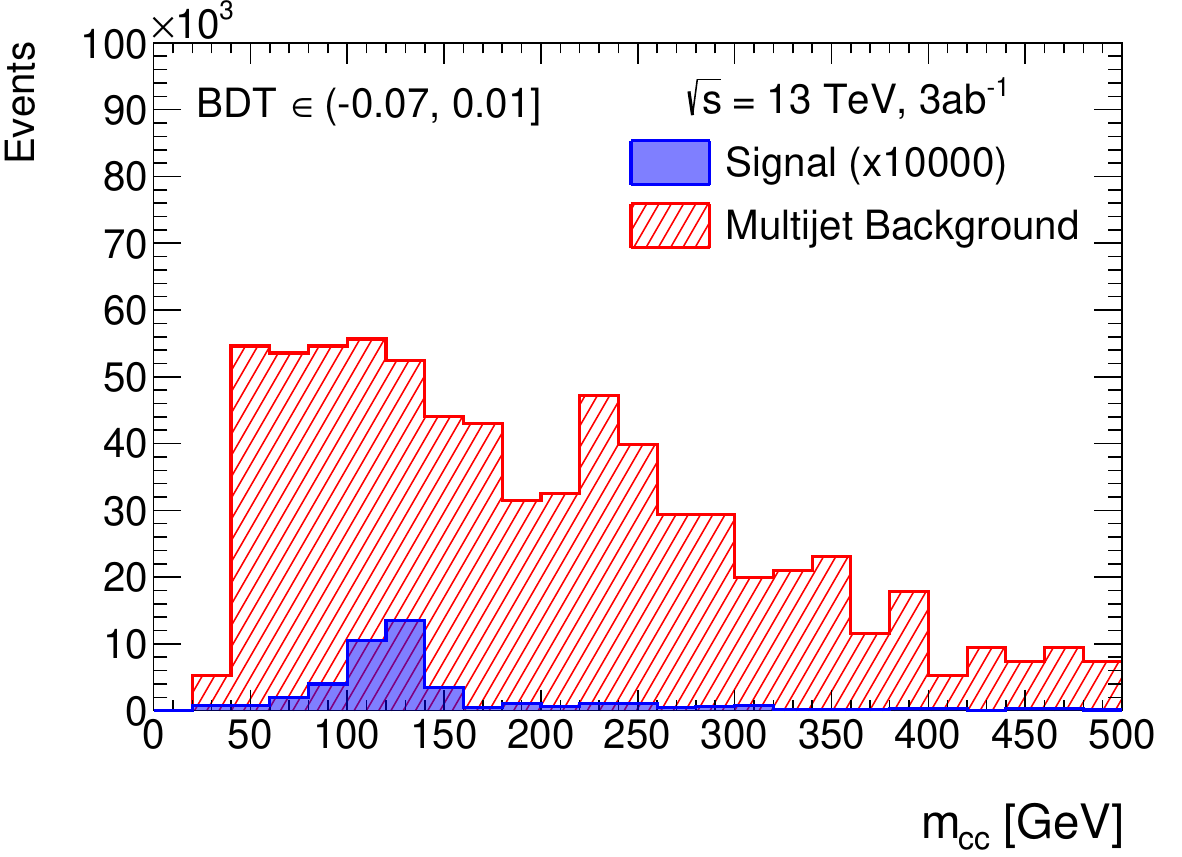}
    \includegraphics[width=0.32\textwidth]{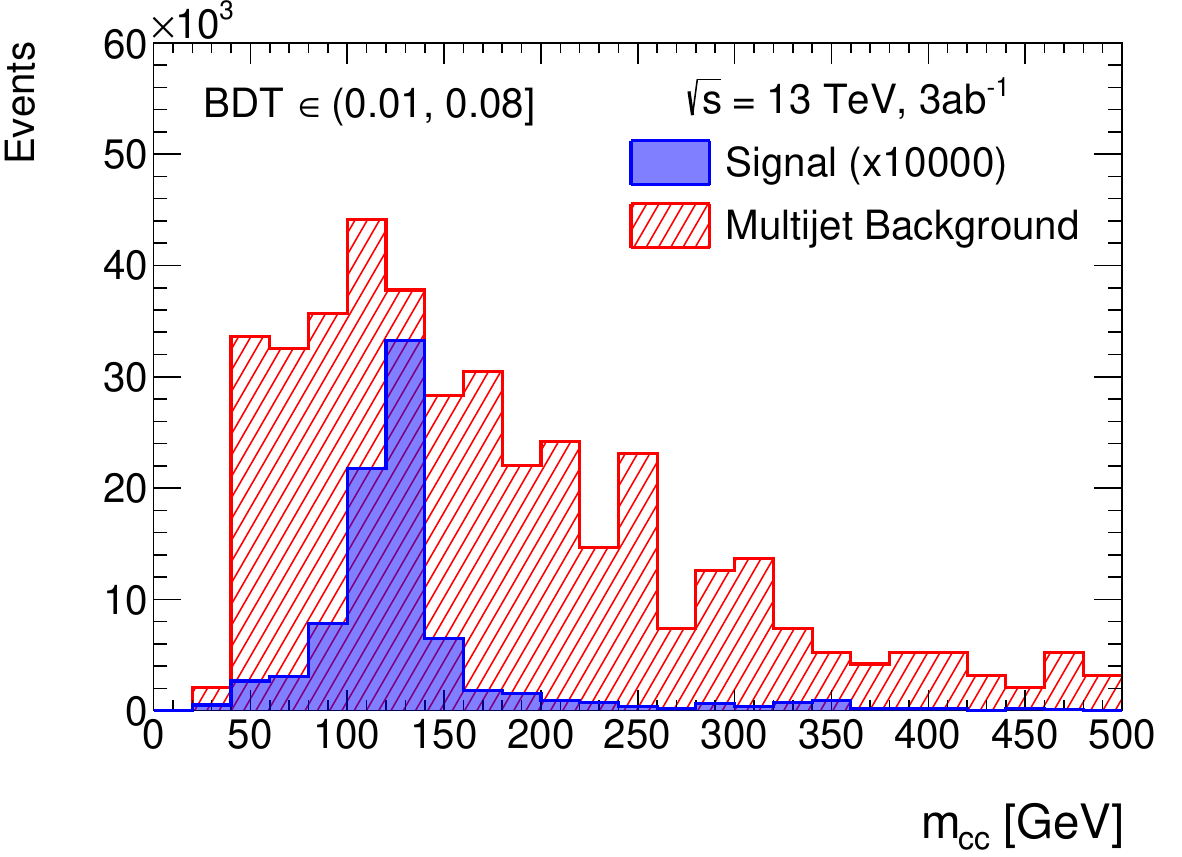}
    \includegraphics[width=0.32\textwidth]{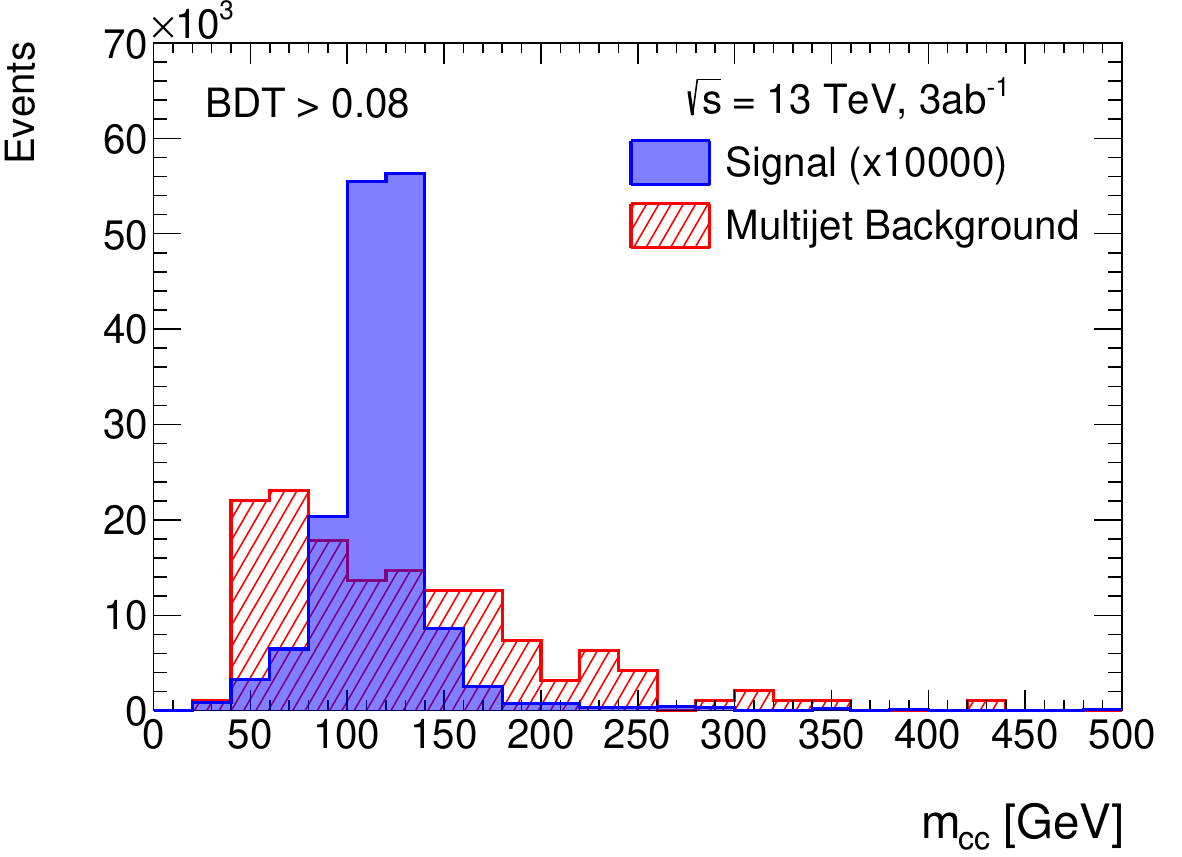}
    \caption{Distribution of the invariant mass of the signal c-jet pair for events in different BDT score intervals, corresponding to the categories: low, medium and high signal.}
    \label{fig:massBDT}
\end{figure*}

To maximize the sensitivity, instead of a single cut, the BDT score is divided into three signal regions: low signal region with BDT score from -0.07 to 0.01, medium signal region with BDT score from 0.01 to 0.08, and high signal region with BDT score $>$ 0.08, as indicated in Table~\ref{tab:bdtnumofeventshl}. The invariant mass distribution of the $c$-jet pair is not used in the BDT training but as a final discriminator and shown in Figure~\ref{fig:massBDT}. The invariant mass distribution $m_{cc}$ could be included in the BDT training to improve the separation between signal and background. However, as is commonly practiced by experiments, we reserved $m_{cc}$ as a most discriminative variable that can be used in a combined fit for signal plus background. A mass window of 100 GeV$-$140 GeV is again selected in the $m_{cc}$ distribution. The expected numbers of events from signal and background of integrated luminosity 3 $\text{ab}^{-1}$ are shown in 
Table~\ref{tab:bdtnumofeventshl} for different BDT score intervals. 
With the BDT cut, we can reach a signal efficiency (background rejection in the interval of BDT score) of 15\% (72\%), 28\% (85\%) and 51\% (95\%) in the low, medium and high signal region respectively. 
In comparison, the signal efficiency (background rejection) of the optimized selection cuts in Sec.~\ref{sec:cut-based} is 24\% (97\%), shown together with the BDT ROC curve on the right panel in Figure~\ref{fig:BDT}. For the same background rejection, BDT can achieve a signal efficiency of 40\%, outperforming the cut-based analysis.
The overall significance is calculated by combining the significance in the three signal regions in quadrature, with the largest contribution coming from the high signal region. A relative change in significance of roughly 50\% is seen, where the relative change is defined as  $(\delta^{\text{BDT}}-\delta^{\text{cut-based}})/\delta^{\text{cut-based}}$, and $\delta = S/\sqrt{B}$.

Like the other extrapolations to the HL-LHC \cite{ATLAS:2019gtt,ATLAS:2018tmw}, we have not considered the effects from the systematic errors. On the one hand, it is important to include the systematic effects to draw robust conclusions, especially given the rather small $S/B$ for the signal searches. On the other hand, the systematic effects due to the background measurement are largely unknown for the HL-LHC. We believe that when the large data sample becomes available, the systematic errors may be controlled to a desirable level of a few percent or lower.

\begin{table}[h]
\centering
\begin{tabular}{c|c|c|c|c|c|c}
\hline
&\multicolumn{2}{c|}{Low signal} & \multicolumn{2}{c|}{Medium signal} &\multicolumn{2}{c}{High signal} \\ \hline
& S & B & S & B & S & B\\ \hline
BDT cut & 4.5 & $7.6\times10^5$ & 8.5 & $4.1\times10^5$ & 16 & $1.5\times10^5$\\ \hline
mass cut Eq.~(\ref{eq:mass})
& 2.4 & $1.1\times10^5$ & 5.5 & $8.2\times10^4$ & 11 & $2.8\times10^4$\\ \hline \hline
$S/\sqrt{B}$ &\multicolumn{2}{c|}{0.0073} & \multicolumn{2}{c|}{0.019} &\multicolumn{2}{c}{0.066} \\ \hline
$S/\sqrt{B}$ combined
& \multicolumn{6}{c}{0.070} \\ \hline
\end{tabular}
\caption{Expected yields of signal and multi-jet background at the HL-LHC with 3 $\text{ab}^{-1}$ from a BDT analysis.}
\label{tab:bdtnumofeventshl}
\end{table}

\subsection{HL-LHC sensitivity to the charm-Yukawa coupling}
\label{sec:HL-LHC}

The expected 95$\%$ $\text{CL}_s$ upper limit on the signal strength $\mu$ in the absence of systematic uncertainties is approximated by $2\sqrt{B}/S$. The BSM modification of the charm-Yukawa coupling is  parametrized using the $\kappa$-scheme as 
\begin{equation}
    y^{\text{BSM}}_c = \kappa_c  y^{\text{SM}}_c,
\end{equation}
then the number of signal events would approximately scale as 
\begin{equation}
    N_{\text{sig}} \simeq \frac{\kappa^2_c}{\kappa^2_H}  N_{\text{sig}}^{\text{SM}},
\end{equation}
where $\kappa_H$ denotes the BSM modification of the Higgs width. In principle, $\kappa_H$ depends on all BSM modifications of SM Higgs decay channels and any new channels. If we assume that $\kappa_c$ is the only non-SM modification, the upper limit on the signal strength can be translated into limit on the charm-Yukawa coupling:
\begin{equation}
    \mu = \frac{\kappa^2_c}{1+Br^{\text{SM}}_{cc}(\kappa_c^2-1)},
\end{equation}
which are shown in Table \ref{tab:kappalimit}, in comparison with some of the existing literature.

\begin{table}[tbh]
\centering
\begin{tabular}{c|c|c||c|c|c|c}
\hline
 & Cut-based & BDT & $ZH$ \cite{ATLAS:2018tmw} & Fit \cite{deBlas:2019rxi} & $Hc$ \cite{Brivio:2015fxa} & $H \rightarrow c\bar{c}\gamma$ \cite{Han:2018juw}\\ \hline
$\mu$  & 43 & 29 & 6.3 & -- & -- & 75\\
    \hline
$\kappa_c$  & -- & 13 & 2.7 & 1.2 & 2.6 - 3.9 & 8.8\\
    \hline
\end{tabular}
\caption{The expected 95$\%$ $\text{CL}_s$ upper limit on the signal strength and the charm-Yukawa coupling from this analysis using 3 $\text{ab}^{-1}$ of data at 13 TeV, respectively, in comparison with other searches as quoted.}
\label{tab:kappalimit}
\end{table}

\subsection{HE-LHC and 100 TeV sensitivity to the charm-Yukawa coupling}

It is natural to ask to what extend the probe to the charm-Yukawa coupling can be improved at the future higher energy hadron colliders, such as the HE-LHC \cite{Abada:2019ono} and the FCC-hh \cite{Benedikt:2018csr}. The answer obviously depends on the detector performance of the charm-tagging, photon detection, and the QCD jet rejection, we nevertheless perform a crude estimate the sensitivity reach by assuming the same detector performance as the HL-LHC study. The sensitivity is estimated by extrapolating the HL-LHC performance as shown in the previous sections. We again calculate the signal cross section and the leading QCD background for $\sqrt{s}$ values of 30 TeV and 100 TeV. By scaling the expected number of signal and background events for $\sqrt{s}$ = 13 TeV to higher energies, we extrapolate the sensitivities, assuming the same luminosity of 3 ab$^{-1}$ as shown in Table~\ref{tab:FCC}.

\begin{table}[h!]
\centering
\begin{tabular}{c|c|c|c}
\hline
$\sqrt{s}$ & 13 TeV & 30 TeV & 100 TeV \\ \hline
$S/\sqrt{B}$ (3 ab$^{-1}$) & 0.07 & 0.14 & 0.25 \\ \hline
$\kappa_{c}$ reach & 13 & 5 & 3 \\ \hline
\end{tabular}
\caption{Expected sensitivity by scaling the collider c.m.~energy $\sqrt{s}$. \label{tab:FCC}}
\end{table}

We have assumed that the cross section increase from $\sqrt{s}$ = 13 TeV to higher energy values does not change as a function of kinematic variables that are used as input to the BDT. Since both signal and background cross sections increase approximately linearly with center-of-mass energy, we see that the sensitivity scales roughly as the square root of center-of-mass energy for the same integrated luminosity, reaching $S/\sqrt{B}$ of 0.25 and $\kappa_c\sim 3$.  

\section{Summary and Conclusions}
\label{sec:conclude}

Testing the charm-Yukawa coupling at the LHC is an important but very  challenging task due to the  overwhelmingly large QCD background. 
In this paper, we first reviewed the existing searches at the LHC and obtained the projection at the HL-LHC in probing the charm-Yukawa coupling, as summarized in Table \ref{tab:summaryofsearches}. 

We proposed to study a new channel: the Higgs boson production via the VBF mechanism plus an additional hard photon in the hope to observe the challenging decay mode $H\to c\bar c$. The additional photon helps for the trigger to this hadronic decay process and to suppress gluon-rich QCD multi-jet background. The search that we proposed can utilize existing ATLAS and CMS triggers or offer new opportunities, for instance utilizing charm tagging in the HLT. We presented our specific proposal for the trigger design in Sec.~\ref{sec:strategy}.

Based on the trigger considerations and the kinematic features of the signal, we first performed a cut-based analysis in Sec.~\ref{sec:cut-based}, which yielded a sensitivity for signal strength $\mu$ of about 43 times the SM value at 95$\%$ $\text{CL}_s$ at the HL-LHC. A boosted decision tree described in  Sec.~\ref{sec:multi} enhanced the sensitivity  by roughly 30$\%$ (using the same definition of relative change as used in Sec.~\ref{sec:multi}), to about 29 times the SM value at 95$\%$ $\text{CL}_s$ at HL-LHC, corresponding to an upper limit of $y_c$ as 13 times the SM value. Our obtained constraint on charm-Yukawa coupling, summarized in Table~\ref{tab:kappalimit}, is better than the $H \to J/\psi + \gamma$ channel~\cite{Aad:2015eyf}.
Even though the limit obtained in our analyses is slightly weaker than the $ZH$ direct search by ATLAS \cite{ATLAS:2018tmw}, our channel will provide complementary information and a combination of different search channels can further improve the limit.

Global analyses of all the Higgs couplings  \cite{Perez:2015aoa,Coyle:2019hvs,deBlas:2019rxi,Carpenter:2016mwd} could result in a more sensitive probe than the direct search result $H\to c\bar c$, but admittedly depending on model-dependent assumptions, such as $|\kappa_V| \leq 1$ etc. Direct measurements of charm-Yukawa coupling are nevertheless indispensable.

Finally, we provided the first investigation of the VBF cross section with an associated photon at higher collider energies of 30 TeV and 100 TeV. Assuming the same signal and background acceptance as well as the similar detector performance, some improvement of the sensitivity would be anticipated, as shown in Table~\ref{tab:kappalimit}. 

As we are entering the new phase of the LHC mission, it is important to push for the challenging measurements and to fully realize the potential for discovery at both the energy frontier and the precision frontier. 

\begin{acknowledgments}
%
This work was supported in part by the U.S.~Department of Energy under grant No.~DE-FG02- 95ER40896, and in part by the PITT PACC. 
\end{acknowledgments}

\bibliography{bibliography.bib}
\end{document}